\documentclass[a4paper,12pt]{article}
\usepackage{amsmath}
\usepackage{gensymb}
\usepackage{graphicx}

\begin{document}

\title{Corner reflectors: fractal analysis and integrated single-photon sources}

\author{Pedro Chamorro-Posada \thanks{e-mail: pedcha@tel.uva.es}\\
  Dpto. de Teor\'{\i}a de la Se\~nal y Comunicaciones\\
 e Ingenier\'{\i}a Telem\'atica,\\
 Universidad de Valladolid, ETSI Telecomunicaci\'on,\\
 Paseo Bel\'en 15, 47011 Valladolid, Spain}

\maketitle
\begin{abstract}
In this work, the properties of the radiation emitted by a corner reflector with an electric dipole feeder are analyzed in the optical domain where the distance between the dipole and the corner apex can be large in terms of the wavelength.  A comprehensive study of the fractal properties of the radiated intensity patterns is presented.  The use of this setup for the realization of single-photon sources in photonic integrated circuits is also put forward and a detailed study of the emission properties of the device and its optimal configurations is presented.  
\end{abstract}

\section{Introduction}

This work focuses on a radiation source composed of a corner reflector and an electric dipole used as a feeder element.  Such an arrangement was put forward by Kraus in 1940 \cite{kraus1940} and soon became an RF antenna of widespread use, and so is nowadays, because it can provide relatively high gain from a compact setup of easy construction.

We address the fractal properties of the fields radiated when the RF setup is translated into the optical domain by wavelength scaling.  Fractal optical patterns, displaying comparable levels of detail across many scales \cite{mandelbrot1982}, can be produced by different means.  In a most straightforward manner, fractal objects modulating the amplitude and/or the phase of optical fields can construct this type of self-similar optical structure.  This is the case of diffractals \cite{berry1979}, fractal lenses \cite{monsoriu2007,machado2017}, fractal photon sieves \cite{gimenez2006}, or fractal zone plates \cite{saavedra2002, ferrando2015}. The generation of optical fractals can also be mediated by nonlinear phenomena, for instance, in soliton fractals \cite{sears2000, yang2000} or the development of temporal \cite{dudley2007,runge2021} and spatial \cite{huang2005} fractal structures under nonlinear propagation. One particularly intuitive mechanism for the build-up of fractal patterns is the accumulation of images in linear optical systems.  This is the case of video feedback systems \cite{courtial2001}, unstable optical resonators \cite{karman1999} or the fractals generated at given distances by the piling of images in the Talbot effect \cite{berry1996}.  At longer wavelengths, in the radio-frequency (RF) region of the electromagnetic spectrum, fractal conductor geometries have also been the subject of intensive research \cite{werner2003,karmakar2020}.  In this case, the use of self-similar layouts typically aims to the implementation of improved performance antennas, either in terms of their size or bandwidth.

One possible mechanism for fractal generation in the device under study could be image accumulation when the bisector angle is drastically decreased. The analyses performed show that fractal structures are not observed in these devices when the distance between the feeder and the corner apex is comparable to the wavelength in RF implementations.  The picture changes as we move the dipole source along the reflector bisector to distances from the corner apex that are large compared to the wavelength.  This results in configurations that can be unfeasible when working in the RF spectral region, but perfectly amenable for optical setups using much shorter wavelengths. It is reasonable to expect similar fractal structures in high directivity antenna systems including reflectors of very large electrical dimensions, like those employed in radio telescopes, satellite earth stations, etc.  On the other hand, very small corner angles are shown to produce highly regular radiation patterns.

A natural arrangement for the implementation of the corner reflector source in the optical regime is obtained by the use of a single-photon atomic or atom-like dipole feeder \cite{toninelli2021,dibos2018,aharonovich2016}.  On a fundamental level, there is a growing interest in the effects of non-integer wavefunction dimensionality in quantum systems \cite{berry1996b} and quantum electronic \cite{kempkes2019} and photonic \cite{biesenthal2022} transport in fractal geometries.  In this context, the proposed setup provides a source of spatial single-photon wavefunctions with a wide range of fractal characteristics.  We also propose the arrangement of an atom-like emitter combined with a corner reflector for the implementation of versatile integrated single-photon sources for quantum information and computation applications and their properties in this framework are further analyzed in this work.

Photons, as flying qubits, are nearly ideal carriers of quantum information \cite{gisin2002,gisin2007,flamini2019}. They also provide a wide range of degrees of freedom for encoding quantum information: time, frequency, polarization, propagation path, optical waveguide modes, orbital angular momentum, etc.  Therefore, single-photon sources are key ingredients of quantum communication systems for quantum key distribution applications and in quantum computer networks.  Even though the mutually non-interacting character of photons is a limiting factor for the implementation of entangling gates that are essential to fully scalable universal quantum computation schemes \cite{divicenzo2000}, single-photon sources are also vital for the broad range of existing photon-based quantum computation schemes \cite{flamini2019}. The simplicity of the corner reflector and its capability for tailoring the radiation properties of an electric dipole emitter in the optical domain, even improving those that have sustained its success in RF over the years, is explored in this work.  In particular, this design is particularly well-suited for its implementation in optical integration platforms providing a highly versatile scheme for single-photon sources \cite{kim2020}.  

The analytical and computational models for unbounded and finite size reflectors used in this work are presented in Section 2.  This is followed by an analysis of the fractal properties of the electrical field radiated at the equatorial plane in each case.    The complexity of the fields produced in the Fraunhofer region by the corner reflector fed by an electric dipole source as we move into the optical regime, both for an ideal and a finite size reflector, are evaluated using the Higuchi fractal dimension \cite{higuchi1988,esteller2001}. The use of corner reflectors with an atom-like dipole feeder for their potential use as single-photon sources for quantum communications and computation is addressed next.  In this latter case, the analysis focuses on the production of highly collimated single-photon spatial wavefunctions.

\section{Analysis of corner reflectors}

\begin{figure}
  \centering
  \includegraphics[width=0.7\textwidth]{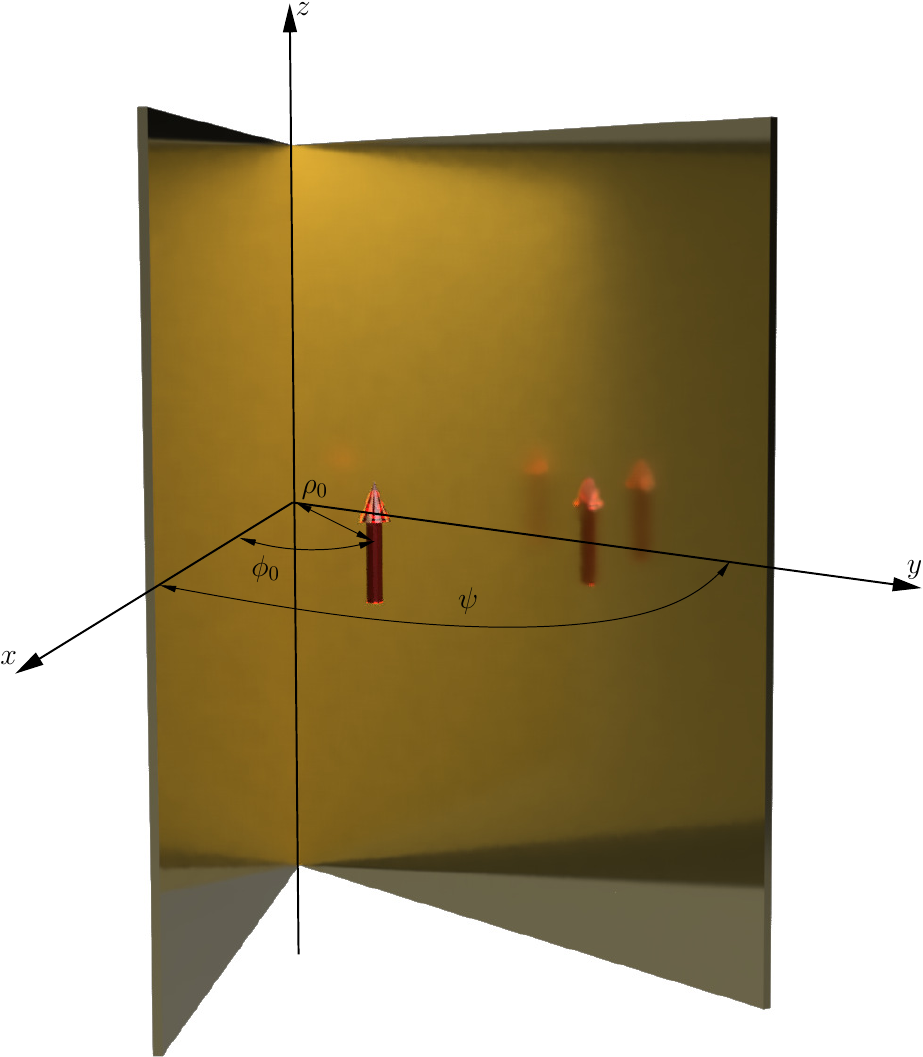}
  \caption{Geometry of the corner reflector with an elementary electric dipole source studied in this work.}\label{fig::geometry}
\end{figure}

Figure \ref{fig::geometry} displays the geometry of the setup that will be analyzed in this work.  $\psi$ is the corner angle limiting a region of space $0\le\phi\le\psi$.   $\left(\rho,\phi,z\right)$ and $(r,\theta,\phi)$ are, respectively, the cylindrical and spherical coordinates of the point where the $\theta$-polarized radiated electric field is observed, and $(\rho_0,\phi_0,0)$ is the position of the electric dipole source which is assumed to be aligned along the $z$ axis.  We will also set the feeder at its most common position at the corner angle bisector: $\phi_0=\psi/2$.

\subsection{Infinite extent reflectors}\label{sec::inf}

We will firstly assume a geometry where the perfect conductor plane mirrors constituting the reflector are of semi-infinite extent.  Under certain conditions, an approximate analysis of the radiation produced by this arrangement can be obtained using image theory \cite{kraus1940} whereby the mirrors are replaced by the equivalent images they produce setting up an equivalent circular array of $2\nu$ electric dipoles when the corner angle is $\psi=180\degree/\nu$, for an integer $\nu$, resulting in an also integer number of current images.  The solution for the general case of non-integer $\nu$ can be approximated by interpolation using the two closest $\nu$ integer solutions \cite{kraus1940}.

The exact solution of the field radiated for the infinite corner reflector was obtained by Wait \cite{wait1954} and Klopfestein \cite{klopfenstein1957} as
\begin{equation}
  E_\theta=\dfrac{j\omega\mu}{\psi}\dfrac{\exp\left(jkr\right)}{r}\sin\left(\theta\right)G(\theta,\phi)\label{eq::sol1}
\end{equation}
where
\begin{equation}
  G(\theta,\phi)=\sum_{n=1}^{+\infty}\exp\left[\left(-j\dfrac{n\pi}{\psi}\right)\dfrac{\pi}{2}\right]\sin\left(\dfrac{n\pi\phi_0}{\psi}\right)\sin\left(\dfrac{n\pi\phi}{\psi}\right)J_{n\pi/\psi}\left(k\rho_0\sin\theta\right).\label{eq::sol2}
  \end{equation}

The electric field strength $E_\theta$ in Eqs. \eqref{eq::sol1} and \eqref{eq::sol2} is defined in the range $0\le\phi\le\psi$, while $E_\theta=0$ otherwise.  As expected for a far field structure, the solution is a spherical wave modulated by the radiation vector $N_\theta=\sin\theta G(\theta,\phi)$ which we will use for the analysis of the radiation patterns. 

We set $\psi=\pi/\nu$, where $\nu$ is an arbitrary real number, and place the source current at $\phi_0=\psi/2$, as discussed above.  We will observe the radiated E-field at the horizontal plane by setting $\theta=\pi/2$.  Using these particular values in \eqref{eq::sol2}, gives
\begin{equation}
  N_\theta=\sum_{n=1}^{\infty}\exp\left(-jn\dfrac{\nu\pi}{2}\right)\sin\left(\dfrac{n\pi}{2}\right) J_{n\nu}\left(k\rho_0\right)\sin\left(n\nu\phi\right)=\sum_{n=1}^{\infty}b_n\sin\left(n\nu\phi\right) \label{eq::expansion}
\end{equation}
which is readily identified as a Fourier series with spectral coefficients
\begin{equation}
b_n=\exp\left(-jn\dfrac{\nu\pi}{2}\right)\sin\left(\dfrac{n\pi}{2}\right) J_{n\nu}\left(k\rho_0\right).  \label{eq::coeficientes}
\end{equation}

The first term in \eqref{eq::coeficientes} is a linear phase associated with the shift the distribution center at the corner bisector from $\phi=0$ to $\phi=\psi/2$.  The second term makes the even harmonics equal to zero and sets a $\pi$ phase shift between consecutive non-null odd harmonics. The last term $J_{n\nu}\left(k\rho_0\right)$ gives the amplitudes of the non-null coefficients.  We note that if the solution is extended out of its definition range $0 \le \phi \le\psi$, for non-integer values of $\nu$, the periodicity of the signal is not consistent with the natural $2\pi$ phase period.

\begin{figure}
  \centering
  \includegraphics[width=0.7\textwidth]{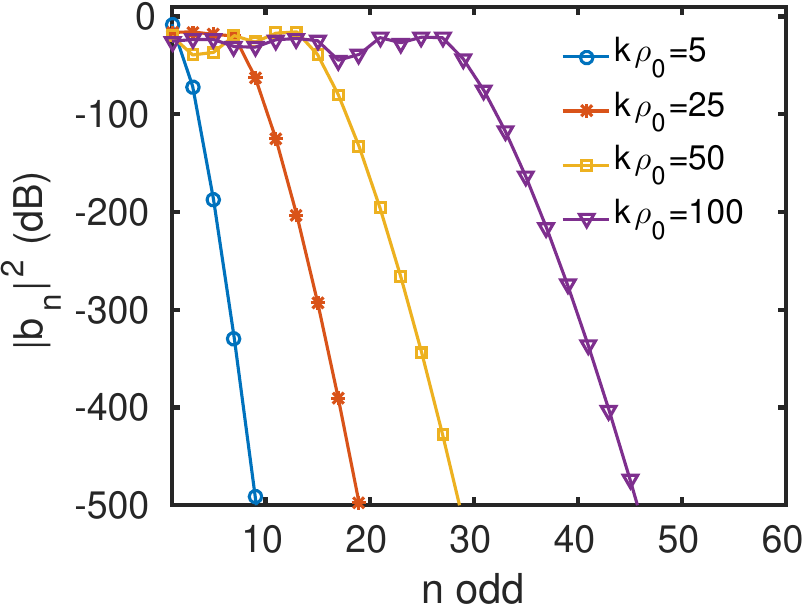}
  \caption{Amplitudes of the odd coefficients of the power angular spectra of the field radiated by the corner reflector for $\nu=3.73$ and four values of $k\rho_0$}\label{fig::espectro}
\end{figure}

The spectral properties of the radiation patterns for a given value of $k\rho_0$  and $\nu$ depend essentially on the amplitude term $J_{n\nu}\left(k\rho_0\right)$ as $n$ varies.  Figure \ref{fig::espectro} displays the power angular spectra as given by the coefficients \eqref{eq::coeficientes} evaluated at odd integers for a particular value of $\nu=3.73$ and four different values of $k\rho_0$.  For all values of $k\rho_0$, we observe an abrupt cutoff in the angular spectrum at a given value of $n=n_{max}$ with a very steep decrease in the amplitudes of the spectral coefficients for $n>n_{max}$. Also, the value of $n_{max}$ increases with $k\rho_0$.  At small values of $k\rho_0$, in the RF regime, where the distance from the corner apex to the source is comparable to the wavelength, there are very few terms contributing to the angular spectra, which results in smooth, very regular, radiation patterns typically found in antenna applications.  It was early observed by Moullin \cite{moullin1945,moullin1949} that the radiated field can be modeled with a negligible error using a single sinusoid for $k\rho_0\le\pi$ and that the error for this assumption is very small even for $\pi\le k\rho_0\le3\pi$.  As the distance between the vertex and the source increases, at the time that we move into the optical domain, the number of  terms contributing to the angular spectra grows.  We expect accordingly increasing complexities in the radiation patterns and, posibly, the full development of self-similar structures as $k\rho_0$ grows. 

\begin{figure}
  \centering
  \includegraphics[width=0.7\textwidth]{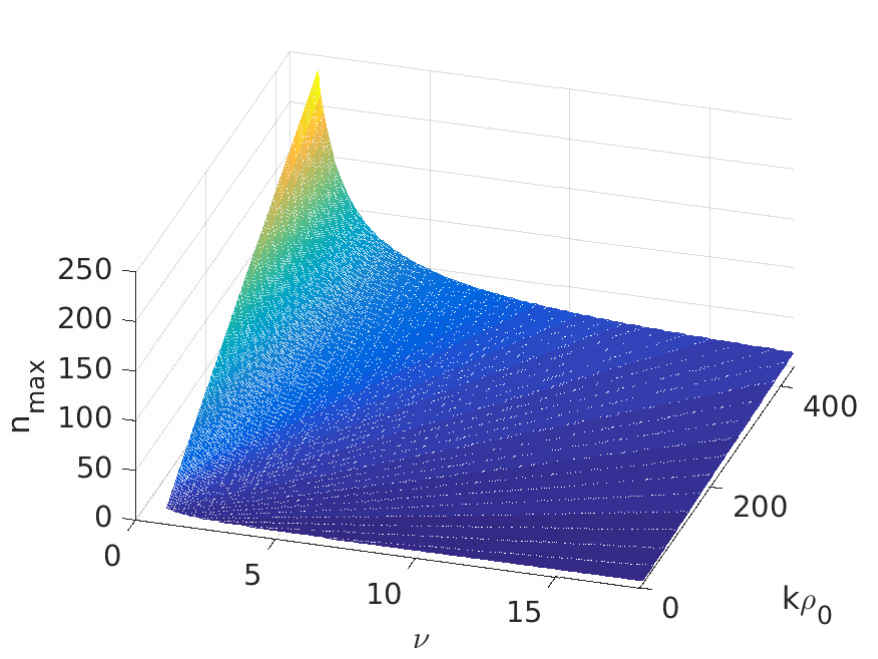}
  \caption{Number of non-negligible terms in the angular spectrum expansion \eqref{eq::expansion} as a function of $\nu$ and $k\rho_0$.}\label{fig::terminos}
\end{figure}

Figure \ref{fig::terminos} displays the number of non-negligible terms in the expansion  \eqref{eq::expansion} evaluated numerically for values of $\nu$ between $1$ to $18$ and $k\rho_0$ ranging from $10$ to $450$.  The results show a steady increase of $n_{max}$ with $k\rho_0$.  Nevertheless, this increase becomes more pronounced as $\nu$ decreases, where the limiting value of $\nu=1$ corresponds to a flat mirror.  Smaller values of $n_{max}$, for a fixed $k\rho_0$, are found as the corner angle narrows with growing $\nu$.

\subsection{Finite size corner reflectors}\label{sec::finite}

To address the radiation produced by finite size corner reflectors, the far field patterns have been calculated using the Method of Moments as implemented in the Numerical Electromagnetics Code \cite{nec2}.

\begin{figure}
  \centering
  \begin{tabular}{cc}
    \raisebox{1cm}{\includegraphics[width=0.35\textwidth]{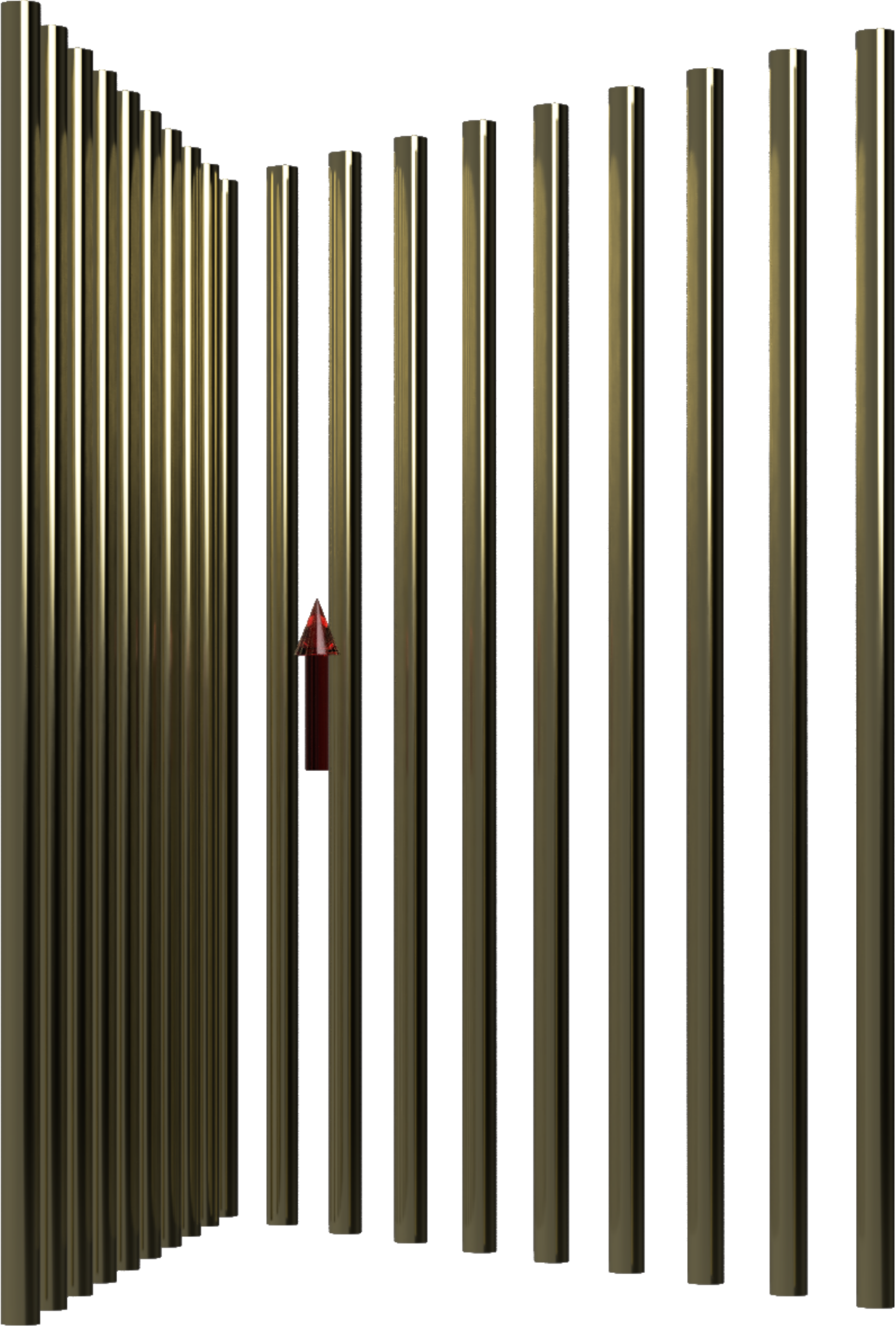}}&\includegraphics[width=0.6\textwidth]{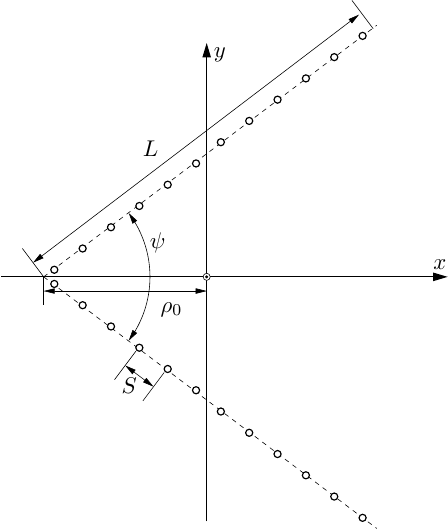}
    \end{tabular}
  \caption{Geometry used for the analysis of finite size mirrors corner reflector where the mirrors have been simulated using conductor rods.}\label{fig::elementos}
  \end{figure}

Practical corner reflectors used in RF antennas usually employ resonant dipoles as feeders.  Nevertheless, we have used as a feeder an electrically short dipole, with a total length that is one tenth of the wavelength, for consistency with the study  based on the exact expressions for an elementary dipole source.  Typical corner antenna designs \cite{arrl} include values of $\nu=1$, $2$, $3$, and $4$, where $\nu=1$ corresponds to a flat mirror.  The value of $\rho_0$ affects the radiation resistance of the antenna, and its typical range of values depends on that of $\nu$, with larger figures employed for the narrower angle reflectors.  All in all, usual RF designs can be found with $0.2\pi\le k\rho_0 \le \pi$.   

As in the typical practical implementations of corner reflector antennas \cite{moullin1945,moullin1949,arrl}, in the numerical calculations, the mirrors have been realized using an arrangement of linear conductive elements as shown in Figure \ref{fig::elementos}.  The diameter used for the elements is $d=0.0083\lambda$ and their height has been set to $H=3\lambda$.  For $k\rho_0>10$, the spacing between elements has been set to $S=0.1 \lambda$ and the total length of the mirrors is the smaller of the two values: $L=2\rho_0/\tan\left(\psi/2\right)$ and $L=30\lambda$.  This bound of the total length has been set to quench the exponential growth of computational resources with the value of $k\rho_0$.  For decreased length reflectors at large values of $k\rho_0$, the central part of each mirror matched to the feeder position, is kept in the setup.  For $k\rho_0\le 10$, the rod spacing is $0.05\lambda$ and the minimal length for the second part of the mirrors at $x>0$ is set to $3\lambda$.

\section{Results}

The results of the survey on the radiation patterns produced by the corner reflector fed by an elementary or small electric dipole are presented in this section.  The analysis of the complexity of the radiation patterns is based on the evaluation of the fractal dimension of the equatorial field intensity distribution.  In this case, we focus on values of corner angles $\psi$ ranging from $30^o$ to $180^o$ where, according to the results of Section \ref{sec::inf}, wider variations in the patterns are expected as $k\rho_0$ is increased.  In the last subsection, we address the analysis of the properties of the radiation emitted for the implementation of single-photon sources.  In this case, we seek highly directive beams emitted along the corner bisector.  Therefore, we include much narrower corner angles in the study, and the radiation field at the full elevation angular range $0\le\theta\le\pi$ is considered.

\subsection{The dipole-apex distance is comparable to the optical wavelength}

\begin{figure}
  \centering
  \begin{tabular}{ccc}
    \includegraphics[width=0.30\textwidth]{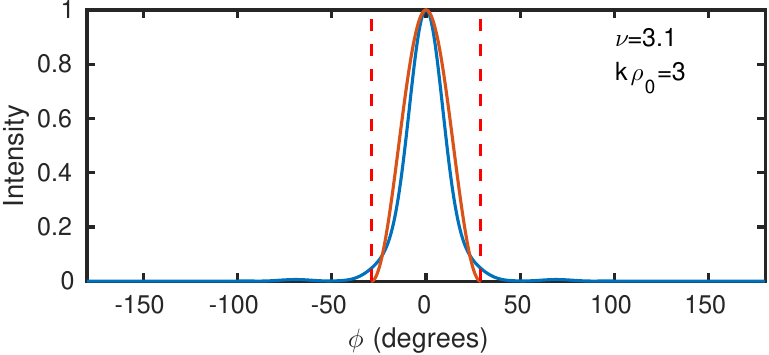}&\includegraphics[width=0.30\textwidth]{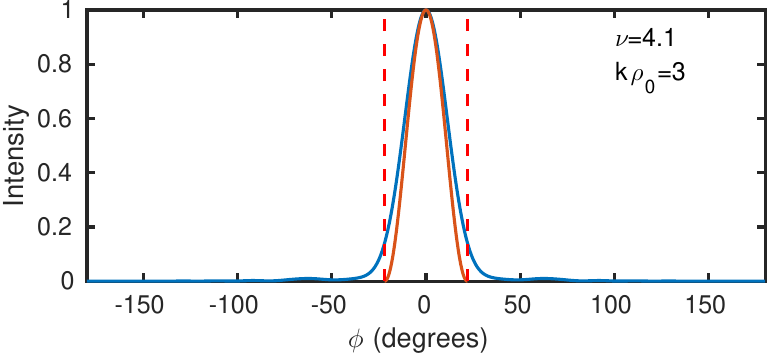}&\includegraphics[width=0.30\textwidth]{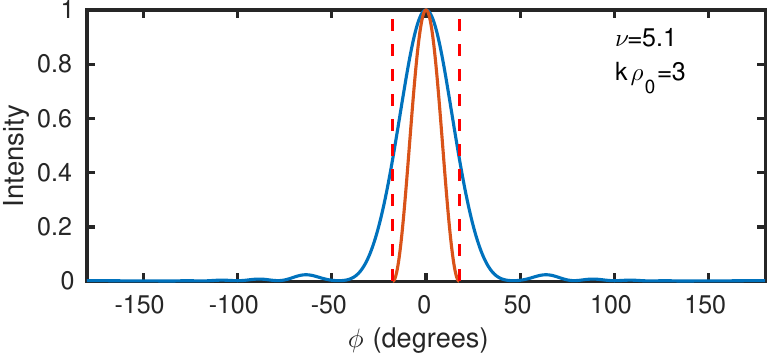}\\
    \includegraphics[width=0.30\textwidth]{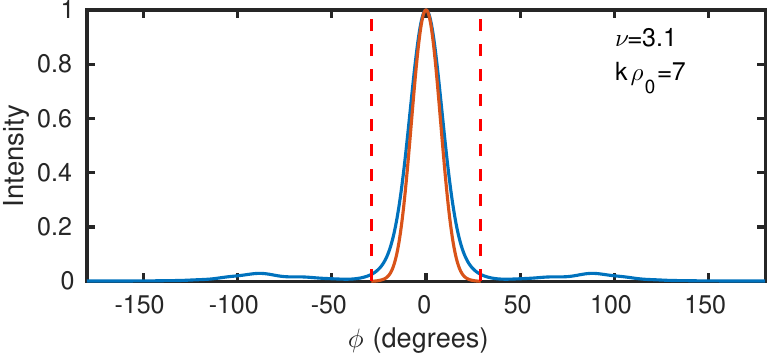}&\includegraphics[width=0.30\textwidth]{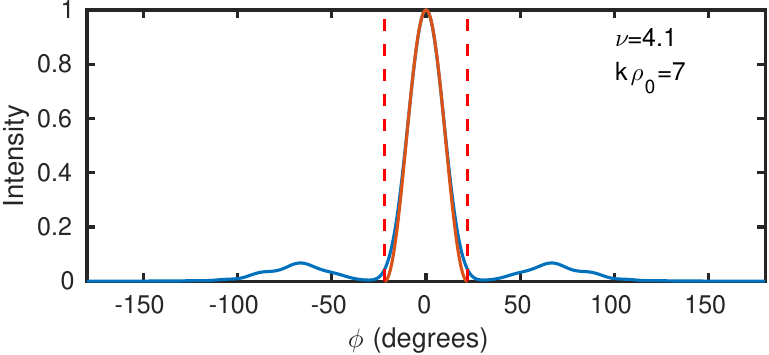}&\includegraphics[width=0.30\textwidth]{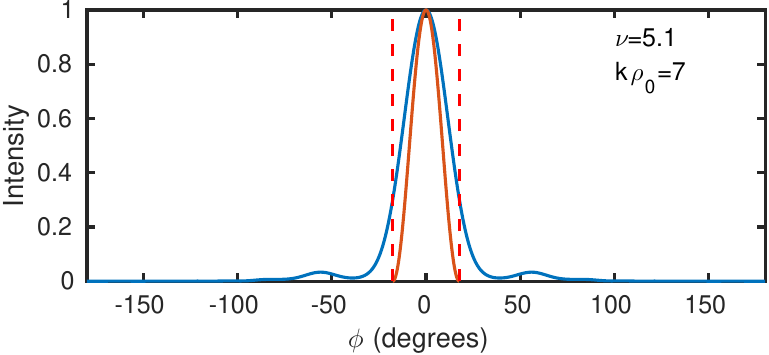}\\
    \includegraphics[width=0.30\textwidth]{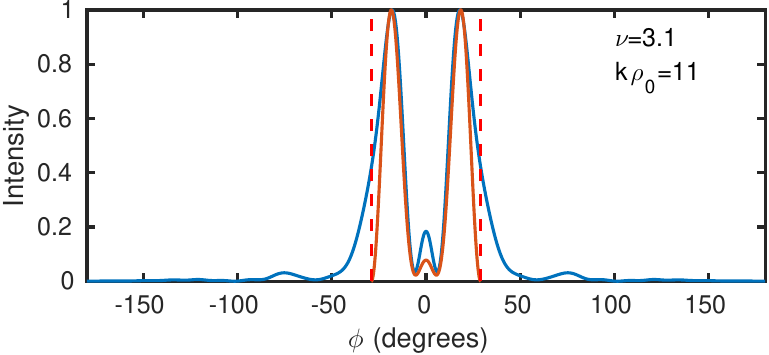}&\includegraphics[width=0.30\textwidth]{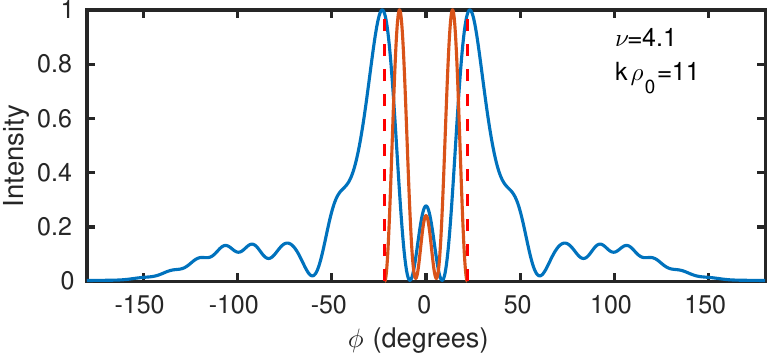}&\includegraphics[width=0.30\textwidth]{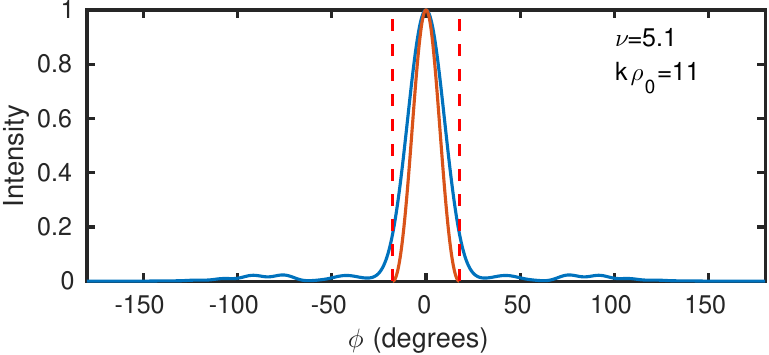}
    \end{tabular}
  \caption{Radiation patterns at small values of $k\rho_0$.  Red lines are for infinite size mirrors and blue for the finite size arrangements.  The red dashed line displays the limits of the corner reflector.}\label{fig::perfiles_p}
  \end{figure}

Figure \ref{fig::perfiles_p} displays the radiation patterns at the equatorial plane when the dipole source is close to the corner apex.   This situation defines the typical RF setup.  The angle $\phi$ has been shifted so the vertex-dipole axis is aligned at $\phi=0$ and the field distributions are centered to facilitate reading the plots.  The results shown correspond to three non-integer values of $\nu$ and $k\rho_0$ equal to $3$, $7$, and $11$.  The plots for infinite and bounded reflectors are traced in red and blue, respectively.   For infinite extent mirrors, the solution is zero for $|\phi|>\psi/\nu$.  The boundaries of this region have been marked with dashed red lines.

There is, in general, a fairly good correspondence between the results obtained in the two cases, even though diffraction effects at the corner edges are evident for the finite size instances.  The radiation patterns shown in Figure \ref{fig::perfiles_p} are smooth, one dimensional curves in agreement with a very small number of terms contributing to their spectral representation \eqref{eq::expansion}.

%
%

\begin{figure}
  \centering
  \begin{tabular}{cc}
    \includegraphics[width=0.45\textwidth]{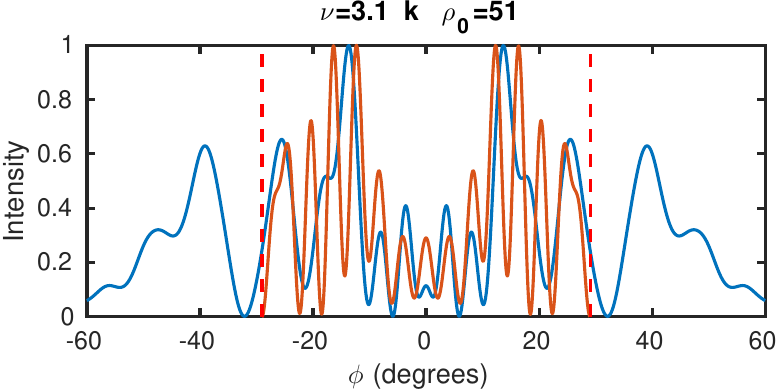}&\includegraphics[width=0.45\textwidth]{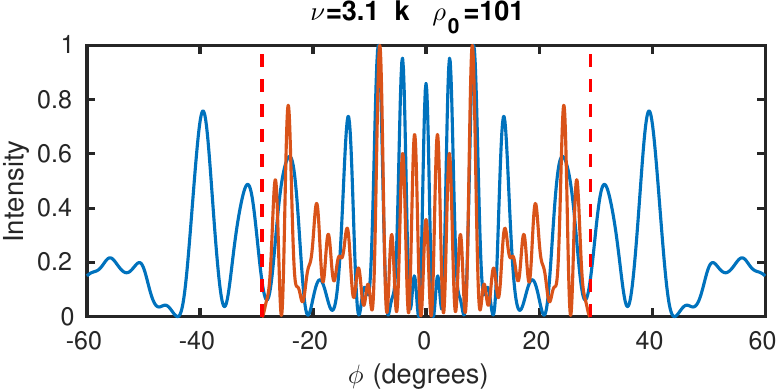}\\
    \includegraphics[width=0.45\textwidth]{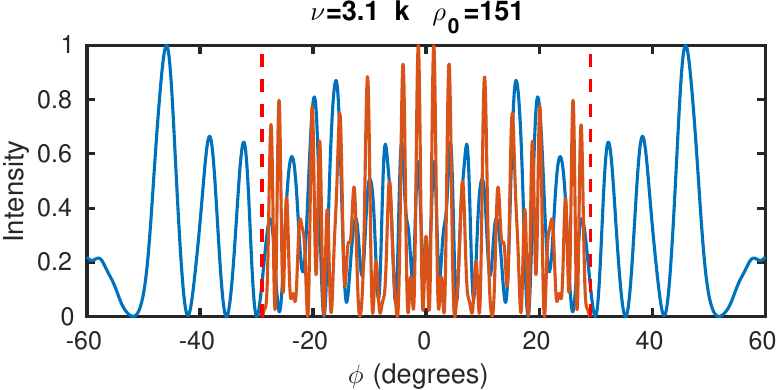}&\includegraphics[width=0.45\textwidth]{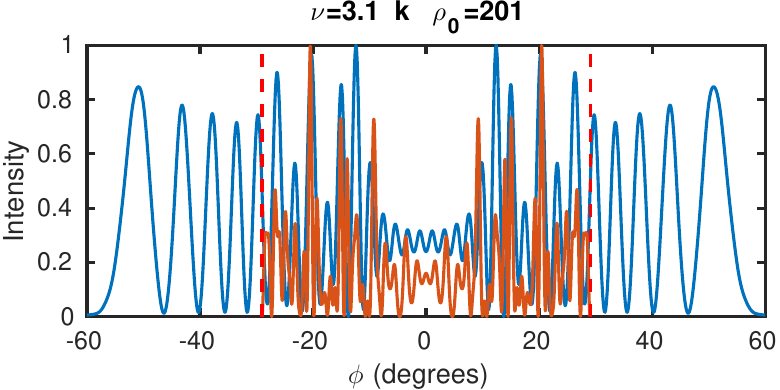}\\
    \includegraphics[width=0.45\textwidth]{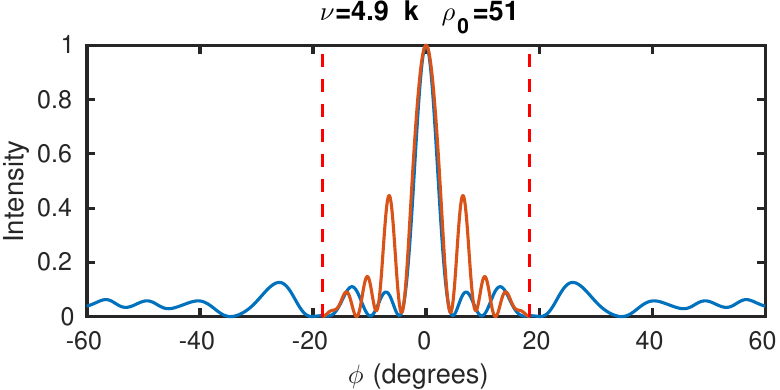}&\includegraphics[width=0.45\textwidth]{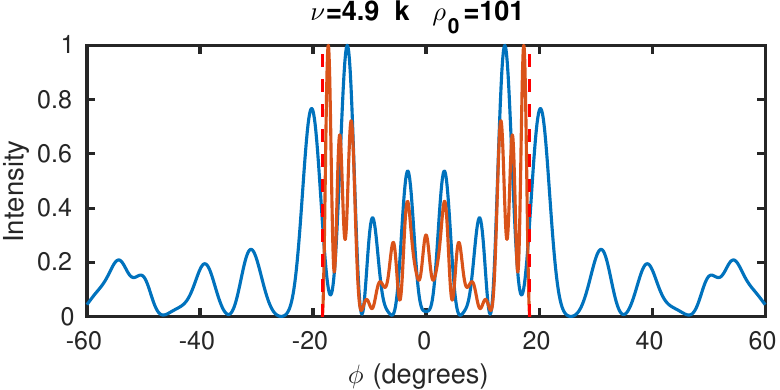}\\
    \includegraphics[width=0.45\textwidth]{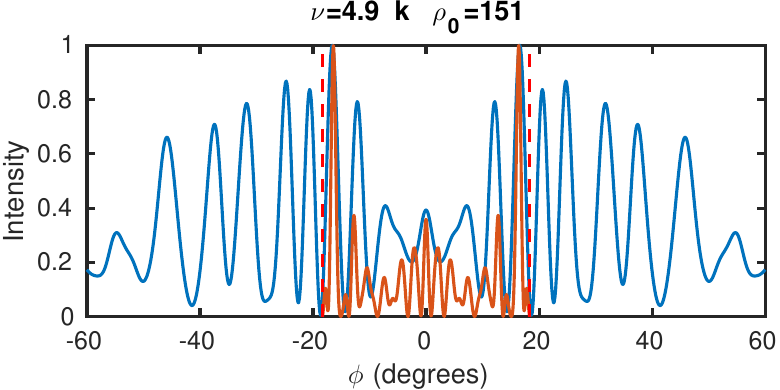}&\includegraphics[width=0.45\textwidth]{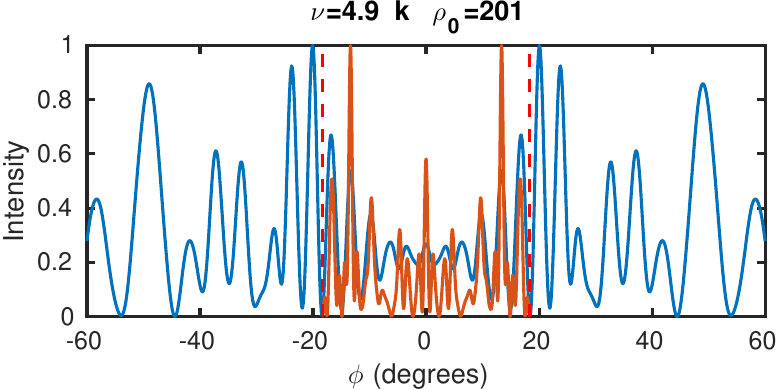}
    \end{tabular}
  \caption{Radiation patterns at larger values of $k\rho_0$.  Red lines are for infinite size mirrors and blue for the finite size arrangement.  The red dashed line displays the limits of the corner reflector.}\label{fig::perfiles_g}
  \end{figure}

\subsection{Moving the dipole feeder further away from the corner cusp: Fractal analysis}

Placing the feeder at arbitrarily large distances, measured in wavelengths, along the corner bisector may be unfeasible for RF implementations, but it is workable when the wavelength is of the order of magnitude of a micron, in the optical domain.

Figure \ref{fig::perfiles_g} compares the radiation patterns obtained both in the finite size case and for unbounded reflectors where the results are traced, again, with blue and red lines, respectively.  Two values of $\nu$, $3.1$ and $4.9$, and four increasing values of $k\rho_0$, $51$, $101$, $151$, and $201$ are displayed. Even though the patterns are, in general, more intricate than those of Figure \ref{fig::perfiles_p},  there is still a reasonable resemblance between the calculations for finite and infinite size setups.  In particular, it can be visually observed a good correlation between the level of intricacy of the patterns in the two cases as the defining parameters are varied.  Correspondingly to the results of section \ref{sec::inf}, the complexity of the patterns increases as $k\rho_0$ grows and/or $\nu$ diminishes, in agreement with the rise in the number of non-negligible terms contributing to the spectral representation of the patterns.

A comprehensive analysis of the complexity of the radiation patterns for a broad region of the parameter space set by $k\rho_0$ and $\nu$ has been performed by calculating their Higuchi fractal dimension \cite{higuchi1988,esteller2001}.  The exploration region extends for $\nu$ ranging from $1$ to $6$ and $k\rho_0$ from $1$ to $500$.  The survey is somewhat more limited in the case of finite size mirrors due to constraints imposed by the computational load. The values analyzed in this later case cover $\nu$ from $3$ to $6$ and $k\rho_0$ from $1$ to $200$.  Modifications of the corner geometry have been introduced in some of the numerical calculations for finite size mirrors to ease the computational load as the mirror sizes grow.  These are detailed in Section \ref{sec::finite}.

\begin{figure}
  \centering
  \includegraphics[width=0.7\textwidth]{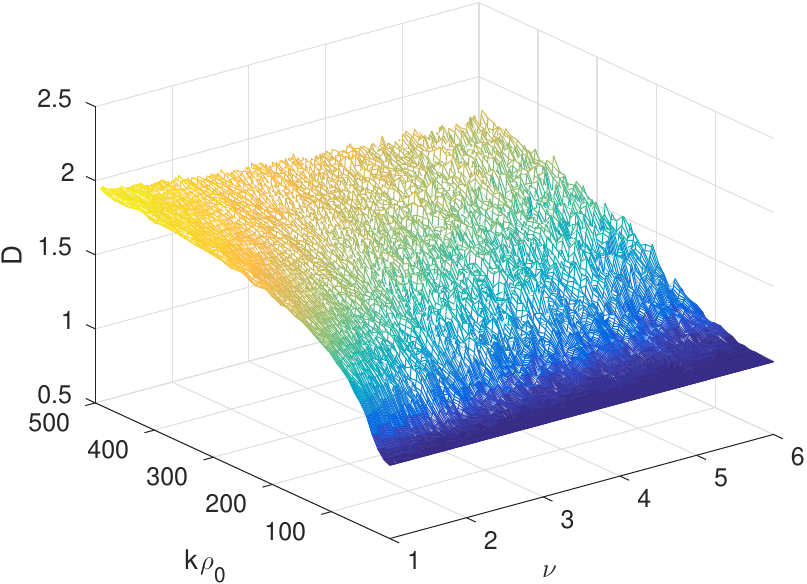}
  \caption{Fractal dimension of the equatorial field distribution produced in the infinite mirror model as a function of $k\rho_0$ and $\nu$.}\label{fig::df-inf}
  \end{figure}

\begin{figure}
  \centering
  \includegraphics[width=0.7\textwidth]{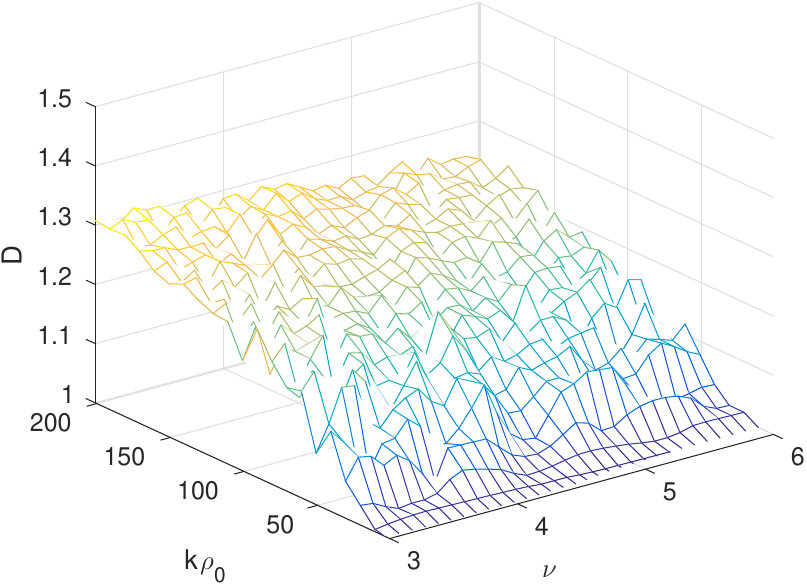}
   \caption{Fractal dimension of the equatorial field distribution produced in the infinite mirror model as a function of $k\rho_0$ and $\nu$.}\label{fig::df-fin}
  \end{figure}

When the distance of the electric dipole feeder to the apex is comparable to the optical wavelength, at small values of $k\rho_0$, the fractal dimensions read  in Figures \ref{fig::df-inf} and \ref{fig::df-fin} are equal or very close to one.  This is the aforementioned RF regime.  As $k\rho_0$ increases when we move deeper into the optical regime, the fractal dimension also grows.  This growth is faster the smaller the value of $\nu$.  This behavior is consistent with the general properties of the power spectrum of the equatorial field depicted in Section \ref{sec::inf}.  For very large values of $k\rho_0=500$ and $\nu=1$, the radiation pattern approaches a space filling curve of fractal dimension equal to 2.

The qualitative behavior of the fractal dimension as a function of $\nu$ and $k\rho_0$ described in the previous paragraph applies both to the results for finite and infinite size reflectors as displayed in Figures \ref{fig::df-inf} and \ref{fig::df-fin}, respectively.  Nevertheless, the smoothing of the patterns associated with diffraction effects at the corner edges of finite size structures produces a reduction in the values of the calculated fractal dimensions.  Whereas the peak fractal dimension observed in Figure \ref{fig::df-inf} is $1.5$, the maximum observed in Figure \ref{fig::df-fin} for the same range of values of $k\rho_0$ and $\nu$ is $1.3$.

An interesting simple case is the limit case of $\nu=1$, corresponding to a single flat mirror extending over the whole $y=0$ plane.  If we assume a dipole source directed in the $z$ direction in the presence of an infinite-extent mirror, the field radiated in the $y>0$ region can be calculated straightforwardly using image theory.  The result so obtained, evaluated at the equatorial plane $\theta=\pi/2$,
\begin{equation}
  \mathbf{N}_\theta=j\dfrac{1}{2}\sin\left(k\rho_0\sin\left(\phi\right)\right)=j\sum_{n=1,n odd}^{\infty}J_n\left(k\rho_0\right)\sin(n\phi)\label{eq::expnu1}
\end{equation}
coincides with that of Equation \eqref{eq::expansion}.  Equation \eqref{eq::expnu1} follows from the equality \cite{abramowitz}.  It is noteworthy that the fractal dimension of the function $\sin\left(k\rho_0\sin(\theta)\right)$ has been formerly addressed in the context of the classification of signal modulation formats based on fractal dimension \cite{elkisky}. The limit case $\nu=1$, also models radio links at frequencies above 50 MHz and propagation distances such that a flat earth is a reasonable approximation.  Signal transmission in this case is mainly affected by the interference of the waves from the emitter antenna and that of its image from the flat earth surface \cite{collin}.  The coverage diagrams that characterize this effect at the receiver end of the link display a very fine lobe structure when the electrical distance of the transmitter to the ground plane $k\rho_0$ is very large, as it happens typically in high frequency links, consonantly with the fractal analysis of the radiation patterns when $\phi$ is close to $180^o$.

\subsection{Narrowing the corner angle for integrated single-photon emitters}

Single-photon sources are pivotal components of emerging photonic quantum information technologies including quantum cryptography \cite{gisin2002}, quantum metrology \cite{giovannetti2011}, and photon-based quantum computing schemes \cite{flamini2019}.  There is solid progress in the development of atom-like single-photon electric dipole emitters, either as isolated organic molecules \cite{toninelli2021}, atoms \cite{dibos2018}, or solid-state sources \cite{aharonovich2016} that can be incorporated into photonic integrated circuits using hybrid integration methods \cite{kim2020}. A crucial requirement of the single-photon waveforms sourced in photonic quantum integrated circuits is their production as highly collimated beams either for free-space applications or for coupling to optical waveguides such as external optical fibers or in-circuit photonic integrated waveguides.  We propose to integrate these elementary single-photon dipole emitters with a corner reflector, taking advantage of the excellent radiation properties of the combination.  High-reflectivity mirrors can be implemented in photonic integrated circuits using metallic coatings \cite{cherchi2015} or dielectric structures \cite{joannopoulos1995}.  In particular, aluminum has been shown to provide high reflectivity and a very good compatibility with the CMOS-based photonic integrated platforms \cite{cherchi2015}.  The compound of a corner reflector and single-photon dipole emitter is a simple and highly versatile photonic integrated single-photon source.  Besides the practical applications in photonic quantum information processing devices, when coupled to free space from the circuit edge, the system can also source photons with quantum wave functions having self-similar properties of more fundamental interest.

One major limitation of atom-like emitters is their typically long radiative lifetimes that hinders the achievement of high repetition rate single-photon emision.  A solution is the enhancement of the spontaneous emission based on the Purcell effect by placing the dipole sources in resonant optical nanostructures such as dielectric cavities or plasmonic nanoantennas.  Designs reported in the literature include dipole-type \cite{bowtie,dipole} and patch \cite{patch} nanoantennas.  At the same time, these elements will shape the polarization state of the emitted light. Lifetime shortening and output coupling of the photon place contradictory requirements on a single-element resonant nanoantenna, whereas the blend of a small spontaneous emission enhancement structure and a large antenna expedites fast photon emission \cite{bogdanov}. We envisage a system designed to tailor the emission properties of atom-like sources in order to achieve highly efficient single-photon sources composed of multiple elements.  We propose the junction of a highly directive corner reflector with a structure for the resonant enhancement of spontaneous emission and the shaping of the polarization state of the dipole source. 

In this section, we address the generation of highly directive single-photon spatial waveforms  in photonic integrated arrangements combining an elementary dipole emitter and a corner reflector.  Implementations with reflector sizes reaching even hundreds of optical wavelengths and such elementary dipole sources are feasible in this scenario.  Therefore, we base our analysis on the analytical expression for an infinite size reflector with an elementary source given in expressions \eqref{eq::sol1} and \eqref{eq::sol2}.  The study is, again, performed for vertical dipole.  The horizontal polarization case, which is relevant for certain implementations, can be analyzed in the same way using the corresponding expression for the radiated field \cite{klopfenstein1957}.  

\begin{figure}
  \centering
  \includegraphics[width=0.8\textwidth]{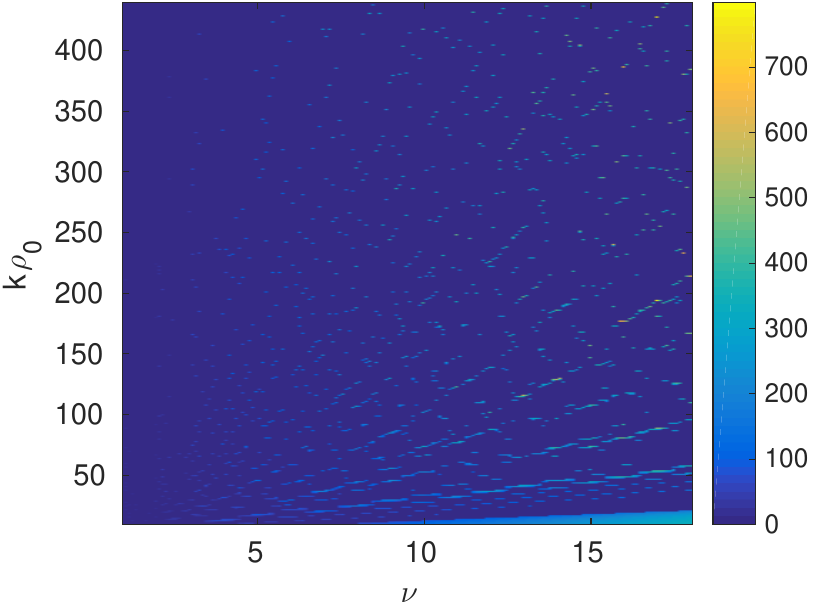}
  \caption{Directivity for the cases when maximal emission takes place along the corner bisector axis.  A null value is assumed otherwise.}\label{fig::D}     
  \end{figure}

 Figure \ref{fig::D} displays the results of an extensive analysis of the radiation properties when the parameters $\nu$ and $k\rho_0$ are varied in the ranges from $1$ to $20$ and from $10$ to $500$, respectively. The figure shows the value of the directivity
\begin{equation}
  D\left(\theta,\phi\right)=\dfrac{4 \pi K(\theta,\phi)}{P},\label{eq::D}
\end{equation}
when the maximum value of $D$ is obtained at  the corner bisector direction $(\theta=\pi/2,\phi=\phi_0=\psi/2)$ and zero otherwise.  In Eq. \eqref{eq::D}, $K$ is the power per unit solid angle (radiation intensity) emitted at each spatial direction $(\theta,\phi)$, and $P$ is the total radiated power.  It is interesting to note how the high directivity values along the corner axis are structured along regions displaying a certain degree of correlation between $\nu$ and $k\rho_0$.

\begin{figure}
  \centering
  \begin{tabular}{cc}
   \hspace*{-2cm} \includegraphics[width=0.7\textwidth]{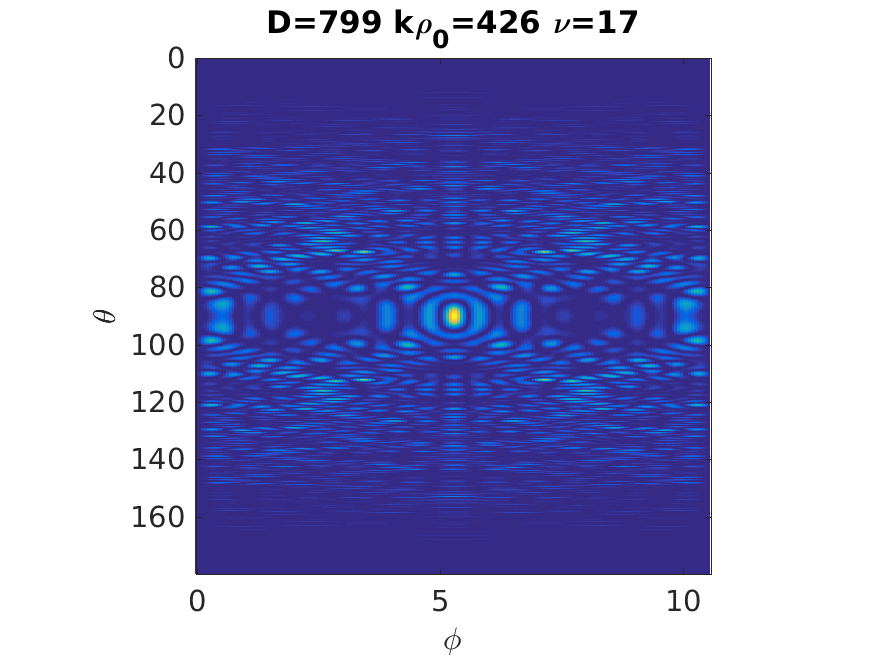}&\hspace*{-2cm}
    \includegraphics[width=0.7\textwidth]{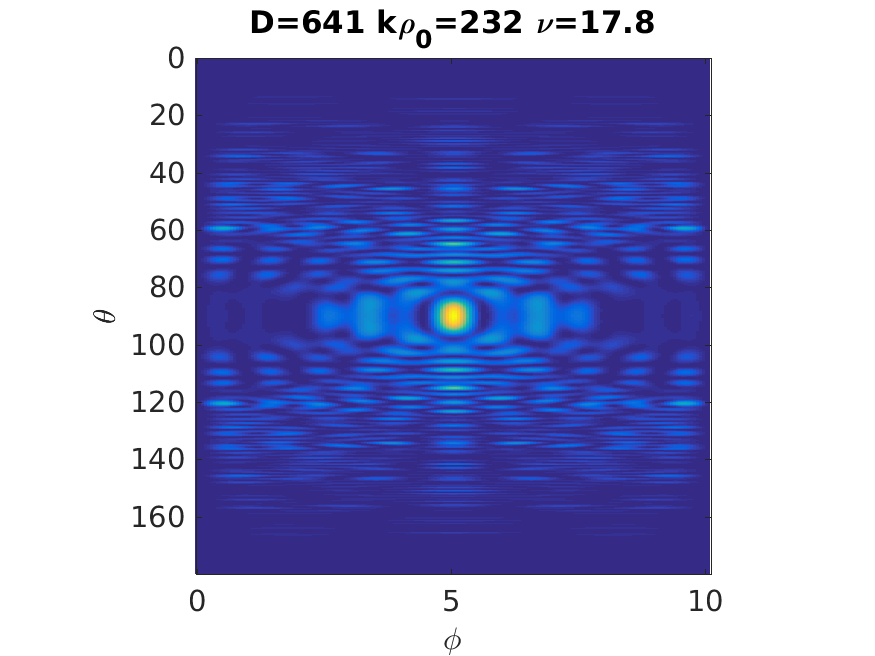}\\
    \hspace*{-2cm}\includegraphics[width=0.7\textwidth]{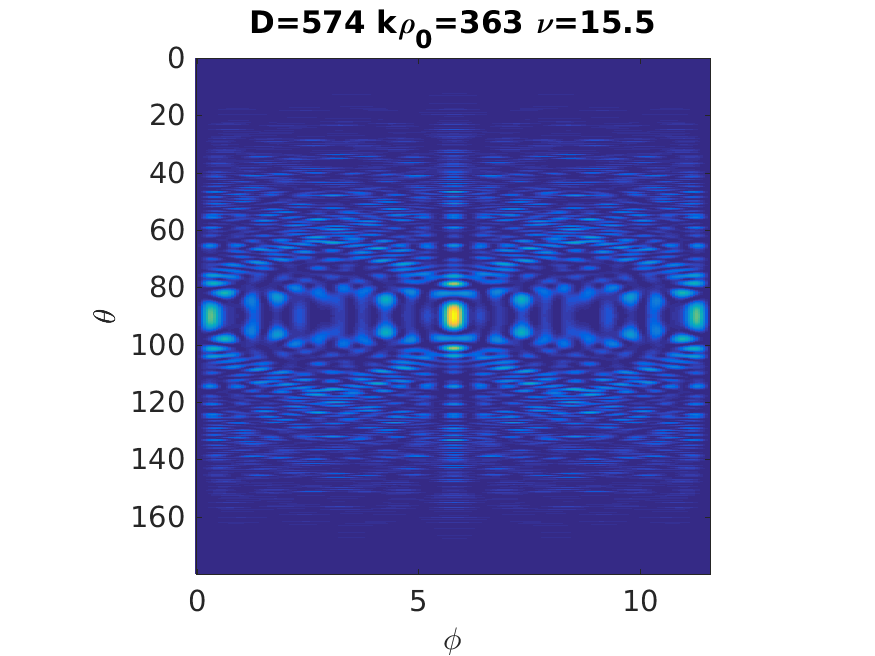}&\hspace*{-2cm}
    \includegraphics[width=0.7\textwidth]{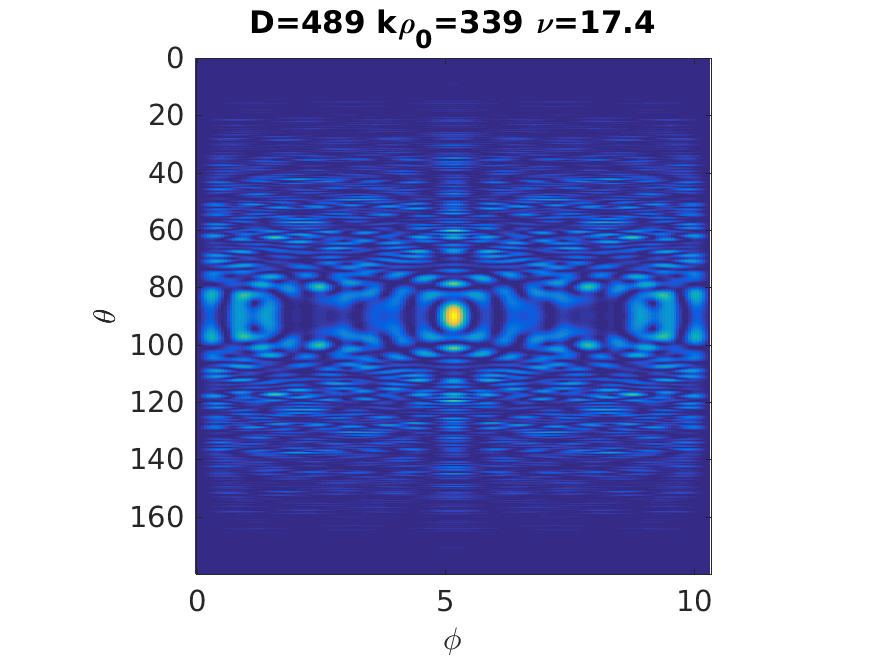}
  \end{tabular}
  \caption{High directivity patterns sampled at large values of $k\rho_0$ and $\nu$. The values of $D$ and those of the parameters $k\rho_0$ and $\nu$ are displayed on top of each figure.}\label{fig::patterns1}
\end{figure}

The greater values of $D$ in Figure \ref{fig::D} are obtained for large values of $k\rho_0$ and $\nu$, in the upper right region of the figure.  In this parameter region, a large value of $k\rho_0$ gives a broader angular spectrum and a larger complexity in the radiation pattern that permits the realization of highly localized structures.  On the other hand, a larger value of $\nu$ reduces the angular extent of the radiation, facilitating obtaining large values of the directivity.  Four of the highest directivity patterns sampled in this parameter region are displayed in Figure \ref{fig::patterns1}.  They illustrate how, in this parameter range, it is possible to obtain very high values of the directivity. It is important to note that the horizontal axes of all the patterns displayed in this subsection are not equally scaled.  Instead, the full azimuthal angular extent defined by the value of $\nu$ is used in each case to facilitate the observation of the fine structure of the plots.  Therefore, for a fair comparison of all the cases, it is important to read carefully the range of $\phi$ in the plots.

\begin{figure}
  \centering
  \begin{tabular}{cc}    
    \hspace*{-2cm}  \includegraphics[width=0.7\textwidth]{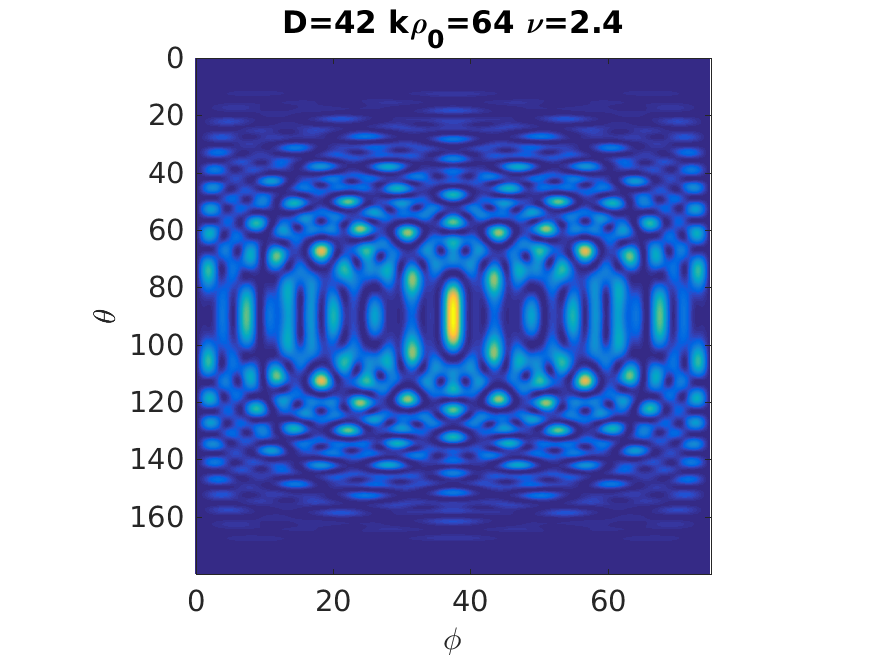}& \hspace*{-2cm}\includegraphics[width=0.7\textwidth]{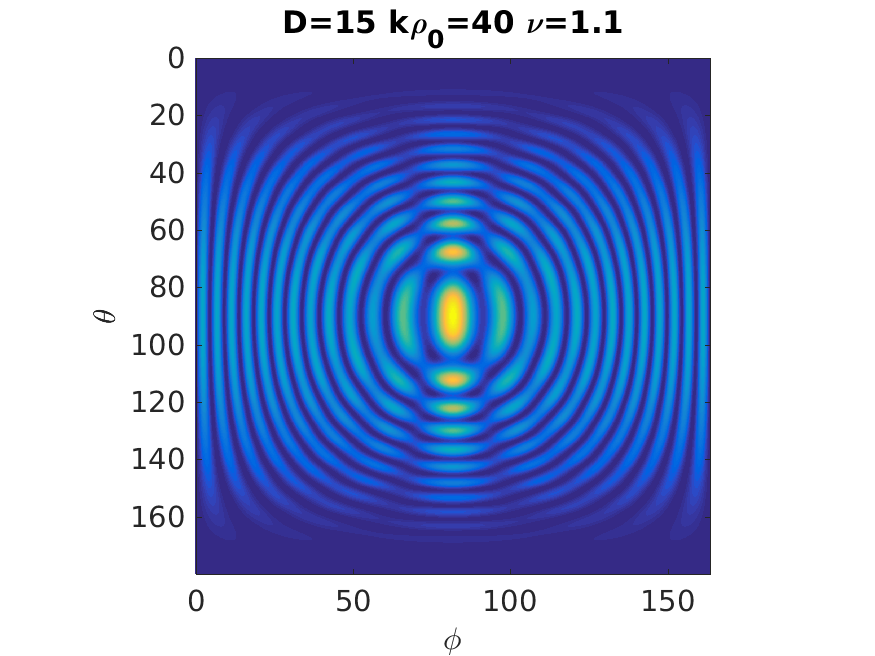}
  \end{tabular}
   \caption{Patterns sampled at small values of $\nu$ and moderate values of $k\rho_0$. The values of $D$ and those of the parameters $k\rho_0$ and $\nu$ are displayed on top of each figure.}\label{fig::patterns2}
\end{figure}

It can also be observed in Figure \ref{fig::D} that, as $\nu$ decreases and the corner angle widens, the left side of the plot becomes depopulated from high directivity on-axis solutions.   Cases with relatively high values of $D$, although one order of magnitude smaller than those shown in Figure \ref{fig::patterns1}, can be found for small values of $\nu$ when $k\rho_0$ is not too small.  Two of these patterns are displayed in Figure \ref{fig::patterns2}. 

\begin{figure}
  \centering
  {\includegraphics[width=0.8\textwidth]{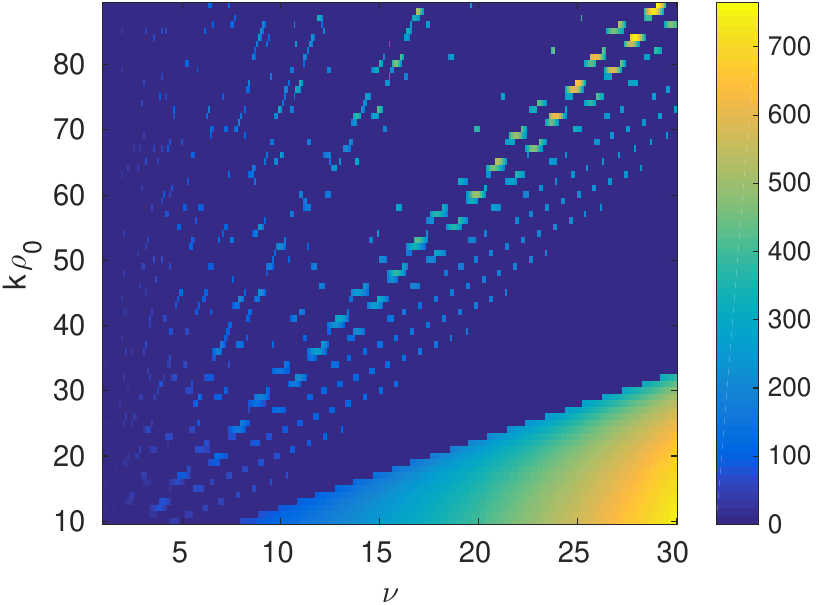}}
  \caption{Directivity for the cases when maximal emission takes place along the corner bisector. A null value is assumed otherwise. In relation to Figure \ref{fig::D}, a limited range of the values of $k\rho_0$ is displayed.}\label{fig::D2}
  \end{figure}

A restricted $k\rho_0$ range view of the values of directivity in Figure \ref{fig::D} is displayed in Figure \ref{fig::D2}.  The aforementioned existing regularities in the disposition of the high values of $D$ can be better observed in this plot.  Also, it is shown how, on the small scale, large values $D$ are found in island-like sets in $(k\rho_0,\nu)$ space.  This localization of the solutions will set a limitation in terms of the fabrication tolerances in the realization of the single-photon source in a photonics integration platform.

Solutions with a large value of $D$ and without restrictive localization features can be observed on the in the broad region in the rightmost part of the bottom of Figure \ref{fig::D2}.  This region corresponds to large values of $\nu$ (very narrow corners) and relatively small values of $k\rho_0$ and its boundary fits well  to the straight line
\begin{equation}
  k\rho_0(\nu) = 1.031\nu+1.65.\label{eq::zona}
\end{equation}
This is the continuation of the domain for typical RF setups when the corner angle is progressively reduced.  As $\nu$ is increased, the maximum value of $D$ grows and the range of $k\rho$ within the region also raises, defining configurations that are accessible to integrated optics implementations with distances between the corner cusp and the dipole source that are large in terms of the optical wavelength.

\begin{figure}
  \centering
  \begin{tabular}{cc}    
    \hspace*{-2cm}  \includegraphics[width=0.7\textwidth]{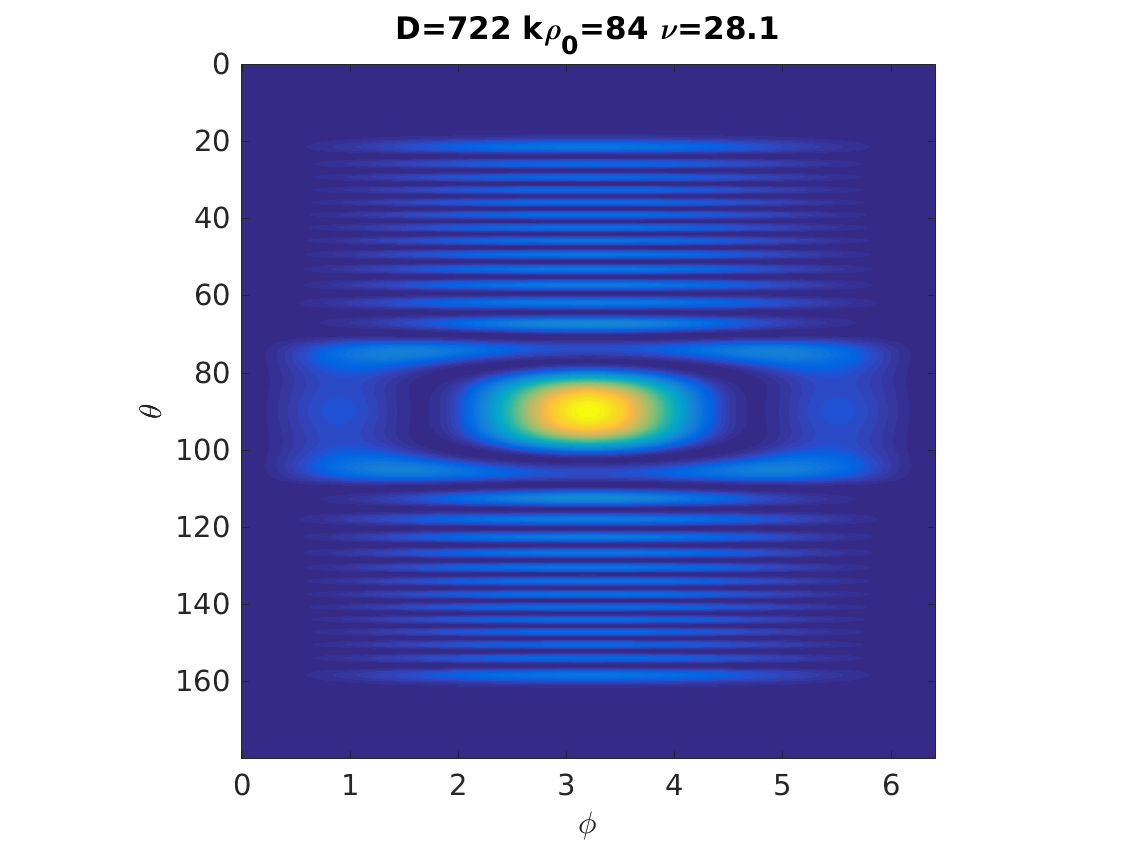}& \hspace*{-2cm}\includegraphics[width=0.7\textwidth]{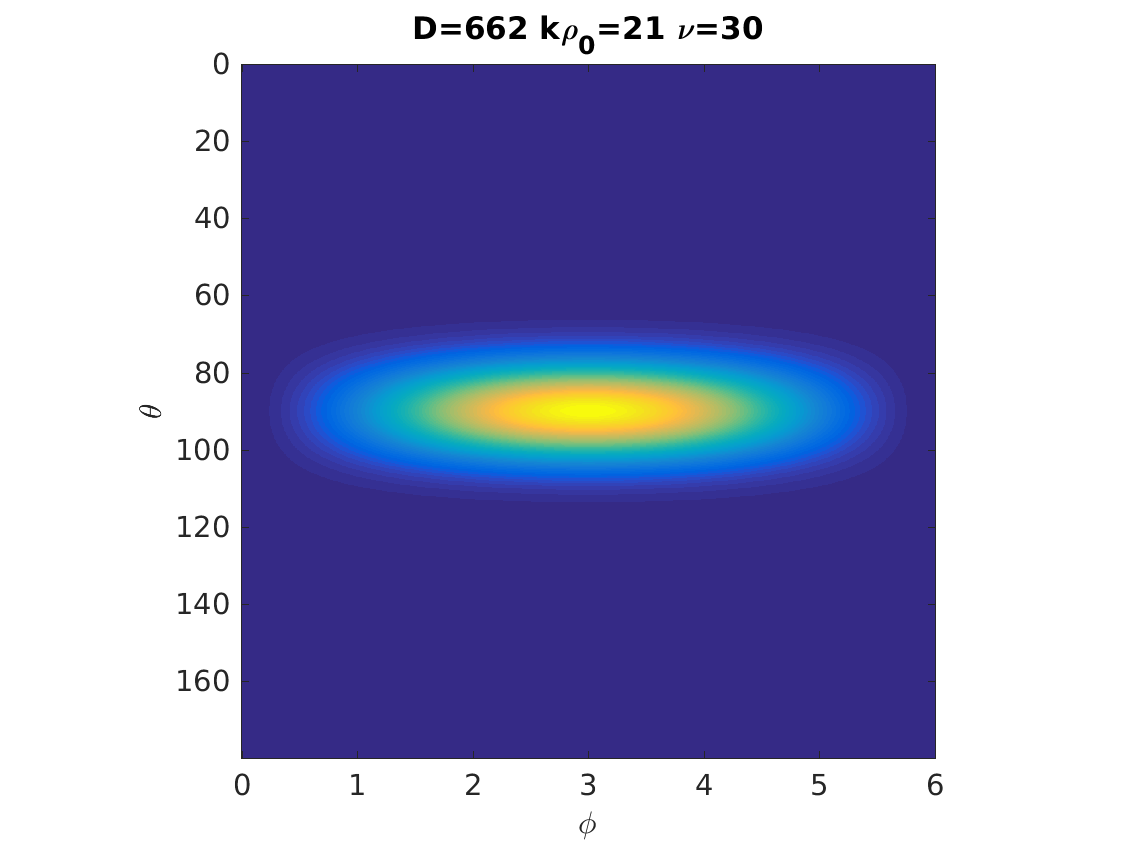}
  \end{tabular}
   \caption{Patterns sampled at the region corresponding to large values of $\nu$ and moderate values of $k\rho_0$ for which the apex-dipole distance is still large in terms of the optical wavelength. The values of $D$ and those of the parameters $k\rho_0$ and $\nu$ are displayed on top of each figure.}\label{fig::patterns3}
\end{figure}

Figure \ref{fig::patterns3} depicts two of the patterns with high values of $D$ found for narrow corner angles, i.e. at large values of $\nu$, in Figure \ref{fig::D2}.  The left plot is a typical solution at the isolated island-like domains in Figure \ref{fig::D2} for intermediate values of $k\rho_0$.  The right plot illustrates the features of the radiation fields for values of $\nu$ and $k\rho_0$ in the broad connected region at the right-bottom corner of Figure \ref{fig::D2} where we find smooth intensity distributions.

\section{Discussion}

We have analyzed the properties of optical implementations of radiation emitters constituted by a corner reflector and an elementary electric dipole source.  This configuration is commonly used as a RF antenna typically employing a resonant dipole feeder.  Whereas in RF implementations the electrical distance between the feeder and the corner cusp is limited due to obvious physical constraints, in optical setups, this separation can be made far larger due to the reduced size of the wavelength.

The production of fractal radiation patterns can be observed when the system parameter space is broadened to address optical setups.  We have presented a detailed analysis of the fractal properties of the intensity patterns produced by the corner reflector combined with an elementary electric dipole source.  The results obtained from the exact expression derived under the assumption of infinite size mirrors are consistent with those of a numerical survey for finite size arrangements.  The fractal dimension systematically increases with the distance of the source to the corner apex.   On the other hand, the fractal dimension also increases with the reflector angle up to the limit case of a flat mirror reflector.  This indicates that the fractal features observed in this work can be of a general character and apply to highly directive antennas employing very large reflectors.  

A device implementation for the realization of integrated single-photon sources based on single-photon dipole emitters has also been put forward.  The fractal analysis then applies to the spatial photon wavefunctions at the quantum level, which is of fundamental interest.  Single-photon sources, on the other hand, are key elements in the emerging field of quantum information technologies.  We have performed a detailed analysis of the emission properties of the proposed setup in this context, where the production of single photons as highly collimated beams is the main target.  On-axis emission with very high directivity can indeed be obtained at various parameter space regions in the proposed integrated optics implementations when one can attain much higher values than those typically obtained in RF systems.  The narrow corner angle region where highly directive and regular radiation patterns are available is of special interest because of the absence of limitations due to fabrication tolerances.  The combination of large volume corner reflectors for shaping the radiation emission with small volume resonant nanostructures to enhance the spontaneous emission rate of atom-like emitters seems particularly promising for the realization of integrated single-photon sources.

\section*{Acknowledgment}
This was supported by the Spanish Ministerio de Ciencia e Innovaci\'on (MCIN), under Grant PID2020-119418GB-I00 and by the European Union NextGenerationEU
under Grant PRTRC17.I1, and by the Consejer\'{\i}a de Educaci\'on, Junta de Castilla y Le\'on, through QCAYLE Project.

\end{document}